\documentclass[preprint2]{aastex61}
\usepackage{amsmath}
\received{}
\revised{}
\accepted{ApJ}

\shorttitle{Evidence for He I 10830 \AA\, in the atmosphere of GJ 3470b}
\shortauthors{J. P. Ninan et al.}
\newcommand{\unit}[1]{\ensuremath{\, \mathrm{#1}}}
\begin{document}

\title{Evidence for He I 10830 \AA~ absorption during the transit of a warm Neptune around the M-dwarf GJ 3470 with the Habitable-zone Planet Finder
}  
\correspondingauthor{Joe P.\ Ninan}
\email{jpn23@psu.edu}

\author[0000-0001-6160-5888]{Joe P.\ Ninan}
\affil{Department of Astronomy \& Astrophysics,  525 Davey Laboratory, The Pennsylvania State University,  University Park, PA, 16802, USA}
\affil{Center for Exoplanets and Habitable Worlds,  525 Davey Laboratory, The Pennsylvania State University,University Park, PA, 16802, USA}

\author{Gudmundur Stefansson}
\affil{Department of Astronomy \& Astrophysics,  525 Davey Laboratory, The Pennsylvania State University,  University Park, PA, 16802, USA}
\affil{Center for Exoplanets and Habitable Worlds,  525 Davey Laboratory, The Pennsylvania State University,University Park, PA, 16802, USA}
\affil{NASA Earth and Space Science Fellow}
\affil{Department of Astrophysical Sciences, Princeton University, 4 Ivy Lane, Princeton, NJ 08540, USA}
\affil{Henry Norris Russell Fellow}

\author{Suvrath Mahadevan}
\affil{Department of Astronomy \& Astrophysics,  525 Davey Laboratory, The Pennsylvania State University,  University Park, PA, 16802, USA}
\affil{Center for Exoplanets and Habitable Worlds,  525 Davey Laboratory, The Pennsylvania State University,University Park, PA, 16802, USA}

\author{Chad Bender}
\affil{Steward Observatory, University of Arizona, Tucson, Arizona 85721, USA}

\author{Paul Robertson}
\affil{Department of Physics and Astronomy, University of California-Irvine, Irvine, California 92697, USA}

\author{Lawrence Ramsey}
\affil{Department of Astronomy \& Astrophysics,  525 Davey Laboratory, The Pennsylvania State University,  University Park, PA, 16802, USA}
\affil{Center for Exoplanets and Habitable Worlds,  525 Davey Laboratory, The Pennsylvania State University,University Park, PA, 16802, USA}

\author{Ryan Terrien}
\affil{Department of Physics and Astronomy, Carleton College, Northfield, Minnesota 55057, USA}

\author{Jason Wright}
\affil{Department of Astronomy \& Astrophysics,  525 Davey Laboratory, The Pennsylvania State University,  University Park, PA, 16802, USA}
\affil{Center for Exoplanets and Habitable Worlds,  525 Davey Laboratory, The Pennsylvania State University,University Park, PA, 16802, USA}
\affil{Penn State Astrobiology Research Center, University Park, PA 16802, USA}

\author{Scott A. Diddams}
\affil{National Institute of Standards and Technology, 325 Broadway, Boulder CO, 80305 USA}
\affil{Department of Physics, University of Colorado, 2000 Colorado Ave, Boulder CO, 80309. USA}

\author{Shubham Kanodia}
\affil{Department of Astronomy \& Astrophysics,  525 Davey Laboratory, The Pennsylvania State University,  University Park, PA, 16802, USA}
\affil{Center for Exoplanets and Habitable Worlds,  525 Davey Laboratory, The Pennsylvania State University,University Park, PA, 16802, USA}

\author{William Cochran}
\affil{Department of Astronomy and McDonald Observatory, University of Texas at Austin, 2515 Speedway, Stop C1400, Austin, TX 78712, USA}

\author{Michael Endl}
\affil{Department of Astronomy and McDonald Observatory, University of Texas at Austin, 2515 Speedway, Stop C1400, Austin, TX 78712, USA}

\author{Eric B. Ford}
\affil{Department of Astronomy \& Astrophysics,  525 Davey Laboratory, The Pennsylvania State University,  University Park, PA, 16802, USA}
\affil{Center for Exoplanets and Habitable Worlds,  525 Davey Laboratory, The Pennsylvania State University,University Park, PA, 16802, USA}
\affil{Institute for CyberScience, The Pennsylvania State University, 525 Davey Laboratory, University Park, PA 16802, USA}

\author{Connor Fredrick}
\affil{National Institute of Standards and Technology, 325 Broadway, Boulder CO, 80305 USA}
\affil{Department of Physics, University of Colorado, 2000 Colorado Ave, Boulder CO, 80309. USA}

\author{Samuel Halverson}
\affil{Jet Propulsion Laboratory, California Institute of Technology, 4800 Oak Grove Drive, Pasadena, CA 91109}

\author{Fred Hearty}
\affil{Department of Astronomy \& Astrophysics,  525 Davey Laboratory, The Pennsylvania State University,  University Park, PA, 16802, USA}
\affil{Center for Exoplanets and Habitable Worlds,  525 Davey Laboratory, The Pennsylvania State University,University Park, PA, 16802, USA}

\author{Jeff Jennings}
\affil{National Institute of Standards and Technology, 325 Broadway, Boulder CO, 80305 USA}
\affil{Department of Physics, University of Colorado, 2000 Colorado Ave, Boulder CO, 80309. USA}

\author{Kyle Kaplan}
\affil{Steward Observatory, University of Arizona, Tucson, Arizona 85721, USA}

\author{Emily Lubar}
\affil{Department of Astronomy and McDonald Observatory, University of Texas at Austin, 2515 Speedway, Stop C1400, Austin, TX 78712, USA}

\author{Andrew J. Metcalf}
\affil{Space vehicles Directorate, Air Force Research Laboratory, 3550 Aberdeen Ave SE, Kirtland AFB, NM 87117 USA}
\affil{National Institute of Standards and Technology, 325 Broadway, Boulder CO, 80305 USA}
\affil{Department of Physics, University of Colorado, 2000 Colorado Ave, Boulder CO, 80309. USA}

\author{Andrew Monson}
\affil{Department of Astronomy \& Astrophysics,  525 Davey Laboratory, The Pennsylvania State University,  University Park, PA, 16802, USA}
\affil{Center for Exoplanets and Habitable Worlds,  525 Davey Laboratory, The Pennsylvania State University,University Park, PA, 16802, USA}

\author{Colin Nitroy}
\affil{Department of Astronomy \& Astrophysics,  525 Davey Laboratory, The Pennsylvania State University,  University Park, PA, 16802, USA}
\affil{Center for Exoplanets and Habitable Worlds,  525 Davey Laboratory, The Pennsylvania State University,University Park, PA, 16802, USA}

\author{Arpita Roy}
\affil{Millikan Prize Postdoctoral Fellow}
\affil{California Institute of Technology, 1200 E California Blvd, Pasadena, California 91125, USA}

\author{Christian Schwab}
\affil{Department of Physics and Astronomy, Macquarie University, Balaclava Road, North Ryde, NSW 2109, Australia}

\begin{abstract}
Understanding the dynamics and kinematics of out-flowing atmospheres of hot and warm 
exoplanets is crucial to understanding the origins and evolutionary history of the exoplanets near the evaporation desert. Recently, ground based measurements of the meta-stable Helium atom's resonant absorption at 10830 \AA~has become a powerful probe of the base environment which is driving the outflow of exoplanet atmospheres. We report evidence for the He I 10830 \AA~in absorption (equivalent width $\sim$ $0.012 \pm 0.002$ \AA) in the exosphere of a warm Neptune orbiting the M-dwarf GJ 3470, during three transits using the Habitable Zone Planet Finder (HPF) near infrared spectrograph. This marks the first reported evidence for He I 10830 \AA\, atmospheric absorption for a planet orbiting an M-dwarf. Our detected absorption is broad and its blueshifted wing extends to -36 km/sec, the largest reported in the literature to date. We modelled the state of Helium atoms in the exosphere of GJ3470b based on assumptions on the UV and X-ray flux of GJ 3470, and found our measurement of flux-weighted column density of meta-stable state Helium $(N_{He^2_3S} = 2.4 \times 10^{10} \mathrm{cm^{-2}})$, derived from our transit observations, to be consistent with model, within its uncertainties. The methodology developed here will be useful to study and constrain the atmospheric outflow models of other exoplanets like GJ 3470b which are near the edge of the evaporation desert.
\end{abstract}

\section{Introduction}
\noindent
The conventional probe for escaping atmospheres has been the Ly$\alpha$ absorption from the ionized exosphere during a planetary transit. This technique has produced exosphere discoveries around hot Jupiters, and hot and warm Neptunes, e.g., HD 209458b \citep{vidalmadjar03}, GJ 436b \citep{kulow14}, and GJ 3470b \citep{bourrier18}. \citet{ehrenreich15} mapped an extended comet-like trail of escaping atmosphere from \object[GJ 436 b]{GJ 436b} using the absorption signatures in the blue wings of Ly$\alpha$. \citet{bourrier18} performed a similar analysis for GJ 3470b using Ly$\alpha$ observations from HST and detected an extended exosphere with neutral hydrogen around GJ 3470b. 

While Ly$\alpha$ is a powerful probe of evaporating atmospheres, it has two major drawbacks. The extinction loss due to interstellar absorption of Ly$\alpha$, as well as the contamination from geocoronal emission, render the central core of the Ly$\alpha$ line unusable. This implies one can probe only the high velocity regions of the exo-spheres via fitting the wings of the Ly$\alpha$ line. The UV observations also have to be done from space---above the Earth's atmosphere---rendering them expensive and hard to do for a large number of systems. 

\citet{oklopcic18} recently suggested the absorption lines of a metastable state of Helium at 10830 \AA~as an alternative probe of evaporating exoplanet atmospheres.  
He I 10830 \AA~lines are not affected by the interstellar medium, and are observable from the ground using high resolution near-infrared spectrographs. Since the core of the line is accessible, even the low velocity base regions of the outflowing exo-sphere is detectable. High resolution spectra enable us to isolate the stellar spectrum from the planet's absorption spectrum which is modulated by the radial velocity of the planet. High resolution is also crucial for removing contamination from narrow telluric absorption lines for ground based observations.  

The search for He I 10830 \AA~absorption is not new. \citet{seager00} proposed that this would be a large signature in F9V type star, HD 209458, though a search by \citet{moutou03} with VLT/ISAAC did not yield a detection. \citet{turner16} had identified He I 10830 \AA~ along with other lines as a potential transition for transit spectroscopy. Recently, a handful of detections were made successfully around K-star planets, namely WASP-107b \citep{spake18,allart19}, HD 189733 b \citep{salz18}, HAT-P-11b \citep{mansfield18, allart18} and WASP-69b \citep{nortmann18}, and one G-star planet HD 209458b \citep{floriano19}.    

In this paper, we report the evidence for He I 10830 \AA\, during the transit of GJ 3470b, a warm Neptune orbiting an M-dwarf star, using the Habitable-zone Planet Finder (HPF) on the 10\,m Hobby-Eberly Telescope (HET) at McDonald Observatory. Section \ref{sec:obs} outlines our HPF observations. We discuss our He I 10830 \AA\, results and associated modelling in Section \ref{sec:results}, and we finally summarize our key conclusions in Section \ref{sec:concl}.

\section{Observations and Data Reduction}
\label{sec:obs}
We observed \object[GJ 3470]{GJ 3470} at different phases of GJ 3470b's orbit using the Habitable-zone Planet Finder (HPF) spectrograph \citep{mah2014SPIE,metcalf19}. HPF is a near-infrared precision radial velocity spectrograph covering the wavelength regime of 8079 -- 12786 \AA~at a resolution of R$\sim$55,000. HPF is actively temperature controlled to the mK level to enable exquisite spectral stability and precision radial velocities in the NIR \citep{stefansson16}. Due to the HET design \citep{HETRamsey, HETqueue}, we are limited to observe for a track length of $\sim$1 hr per night at GJ 3470's declination. We observed three transits of GJ 3470b on 2018-11-30 UT, 2019-01-19 UT, and 2019-04-16 UT, when the transits aligned with the observable window of the HET. Out-of-transit observations were conducted on 2019-01-04 UT, 2019-01-27 UT and 2019-04-17 UT. Three frames of 916 seconds integration time were taken in each of the $\sim$1 hr tracks during in-transit and out-transit observations. The median signal-to-noise ratio of individual spectra was $\sim$100 per pixel for the 1$^{st}$ transit, and $\sim$200 for the 2$^{nd}$ and 3$^{rd}$ transit.  The median number of pixels per resolution element in HPF is 2.8. 

Echelle spectra of GJ 3470 from HPF were reduced using our HPF spectral extraction pipeline. The pipeline first generates 2D flux images (in units of electrons per second) from the up-the-ramp H2RG readout data \citep{ninan18}. A formal pixel-by-pixel variance image is also generated and propagated through the pipeline. The simultaneous Star, Sky and Calibration fiber spectra (as well as variance estimates) are then extracted from this 2D image \citep{kaplan2018}. 

To minimise scattered light contamination, no simultaneous calibration light spectra was used during the observations. The wavelength calibration of the extracted spectra was done using a custom built frequency stabilised laser comb \citep[LFC;][]{metcalf19} measurements taken throughout the night, inversely weighted by the time difference between science and LFC images following the methodology in \cite{metcalf19} and \cite{stefansson20}. All of our analysis and plots are in vacuum wavelengths. The spectra are then flat corrected and deblazed. 

Sky emission lines are subtracted using the simultaneous sky spectrum. This step also subtracts out the smooth background scattered light due to the proximity of the sky fiber to the star fiber image on the detector. A 7 pixel median filter smoothing was applied to the sky fiber data in regions devoid of emission lines to reduce the noise in continuum regions of sky emission spectrum. Stellar continuum was also removed by fitting a quadratic polynomial continuum. Telluric absorption lines were corrected using an improved version of \mbox{TERRASPEC} code \citep{bender2012}, a wrapper around \texttt{LBLRTM} \citep{clough05}. A good sky emission line subtraction and telluric absorption correction was crucial since He~I 10830 \AA\, falls close to a strong OH emission lines as well as water absorption lines. The residuals in the sky and telluric corrected regions are still dominated by imperfect modelling and they are significantly above the photon noise. 

Due to HET's constrained pointing---resulting in short ($\sim$1 hour long) observation windows on GJ 3470---our out-of-transit and in-transit observations are spread across multiple nights. To down-weight the region of the spectrum partially recovered from sky and telluric corrections, the variances of those regions are artificially inflated by a large factor ($\sim$100). In doing so, this allows us to combine multiple epochs (with different barycentric shifts) weighted by the variance and obtain an average spectrum without residual artifacts from imperfect telluric and sky emission subtraction dominating.

The GJ 3470 system's parameters used in this paper are summarized in Table \ref{table:GJ3470properties}.

\begin{table*}[t!]
\renewcommand{\thetable}{\arabic{table}}
\centering
\caption{Orbital parameters of GJ 3470b used in calculation} \label{table:GJ3470properties}
\begin{tabular}{ccll}
\tablewidth{0pt}
\hline
\hline
Parameter & Value & Description & Reference\\
\hline
\decimals
$\gamma$            & 26.090                      & Absolute stellar RV ($\unit{km\:s^{-1}}$) & \citet{gaiaDR218}  \\
$T_{\mathrm{eff}}$  & 3600                        & Stellar effective temperature (K)         & \citet{awiphan16}  \\
$\mathrm{[Fe/H]}$   & 0.20                        & Stellar metallicity                       & \citet{awiphan16}  \\
$\log g$            & 4.695                       & Stellar surface gravity                   & \citet{awiphan16}  \\
$M_*$               & 0.51 $\pm$0.06              & Stellar mass (M$_\odot$)                  & \citet{biddle14}   \\
$R_*$               & 0.48 $\pm$0.04              & Stellar radius (R$_\odot$)                & \citet{biddle14}   \\
$T_0$               & 2456677.727712 $\pm$0.00022 & Transit midpoint ($\mathrm{BJD_{TDB}}$)   & \citet{dragomir15} \\
$T_{14}$            & 0.07992 $\pm$0.001          & Transit duration (days)                   & \citet{dragomir15} \\
$P$                 & 3.3366413 $\pm$0.0000060    & Period (days)                             & \citet{dragomir15} \\
$e$                 & 0.114 $\pm$0.052            & Eccentricity                              & \citet{kosiarek19} \\
$\omega$            & -1.44$^{+0.1}_{-0.04}$      & Argument of periastron (radians)          & \citet{kosiarek19} \\
$K$                 & 8.21$^{+0.47}_{-0.46}$      & RV semi amplitude ($\unit{m\:s^{-1}}$)    & \citet{kosiarek19} \\
$M_{pl}$            & 12.58$^{+1.31}_{-1.28}$     & Mass of the planet (M$_\oplus$)           & \citet{kosiarek19} \\
$b$                 & 0.47$^{+0.074}_{-0.110}$    & Impact Parameter                          & \citet{dragomir15} \\
\hline
\end{tabular}
\end{table*}

\section{Results and Discussion}
\label{sec:results}

\subsection{Evidence for He I 10830~\AA~ Absorption during transits}
\begin{figure*}[t!]
\begin{center}
\includegraphics[width=0.9\textwidth]{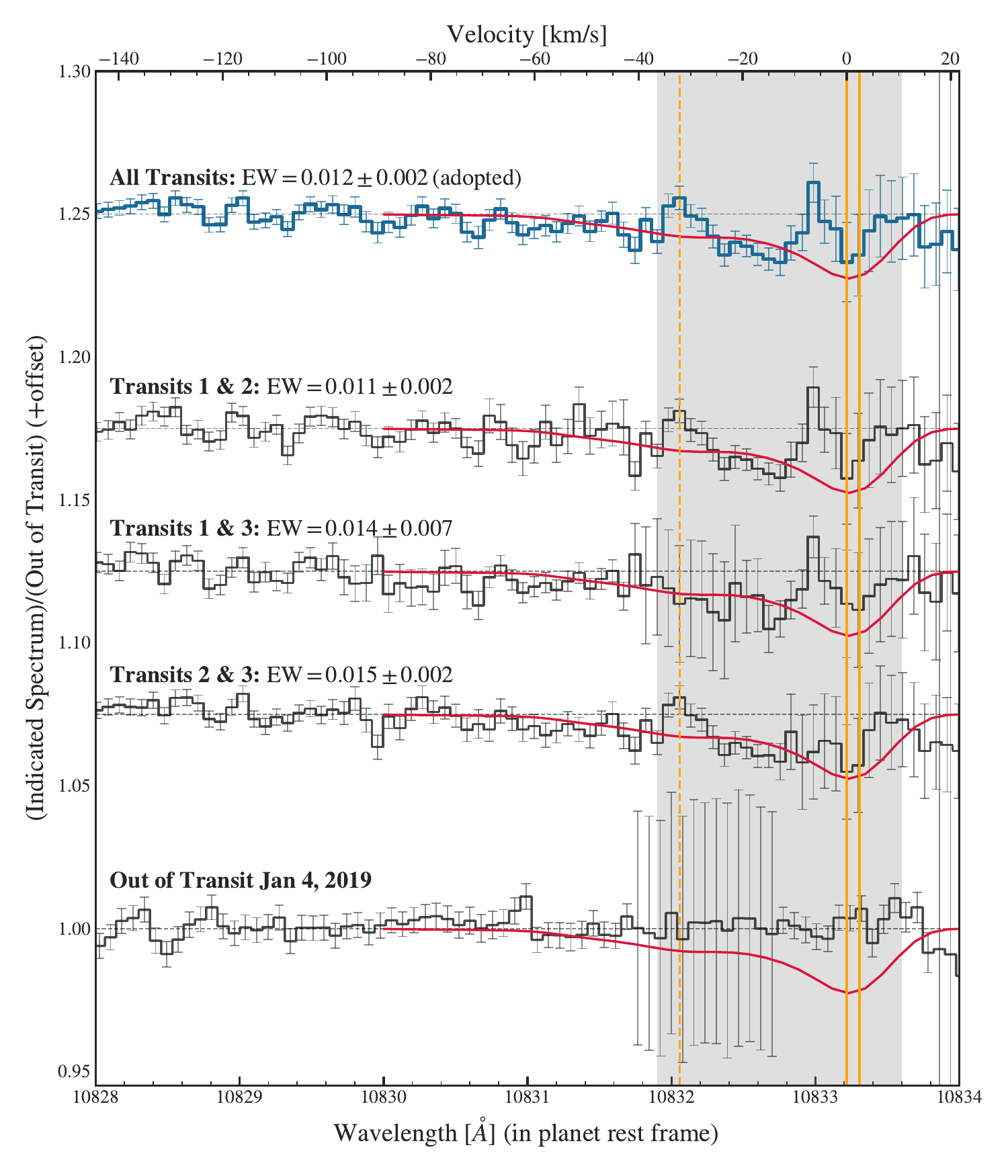}
\caption{ Evidence of broad He 10830 absorption (EW: 0.012 $\pm$0.002\AA) in the wavelength range 10832.2 to 10833.4 \AA~in the in-transit / out-transit spectrum during three transits of GJ 3470b. The x-axis shows vacuum wavelength in planet's rest frame at mid-transit. The rest vacuum wavelengths of the He 10830 \AA\,  triplet lines in planet's rest frame are marked by the vertical dash and solid orange lines. The grey region highlights the broad absorption discussed in this paper, and it is the wavelength window we used to measure reported equivalent widths. The expected line profile from our simple model described in Section \ref{sec:LineProfileModel} is also shown for comparison by the red curves. An out-of-transit epoch divided by the reference out-of-transit spectrum is also shown at the bottom, demonstrating the lack of any absorption in out-of-transit data.} 
\label{fig:GJ3470InByOut}
\end{center}
\end{figure*}

The blue curve at the top of Figure \ref{fig:GJ3470InByOut} shows the weighted average of all the in-transit spectra from three transit epochs divided by the average of the out-of-transit spectra. The vacuum wavelengths of the He 10830 \AA~triplet in the planet's rest frame are marked by the orange vertical lines. Since the telluric correction as well as the sky emission line subtraction are not perfect, regions of sky/telluric contamination have enlarged error bars as discussed in Section \ref{sec:obs}. The weighted average of all combinations of transits taken two at a time is also shown in the curves below. The single peak observed at 10833 \AA~is not present in Transit 2 and 3, it is only seen in Transit 1. However, since 10833 \AA~was separated from the telluric band only during Transit 1, it did not get averaged out in the final weighted average. 
 Since Transit 1 had only half the S/N compared to other two transits we suspect this is an artifact due to photon and detector noise (see blue curves in Figure \ref{fig:IndividualRatioSpectrum} for the individual ratio spectrum from Transit 1). For a null result comparison, one of the out-of-transit epochs is also shown in Figure \ref{fig:GJ3470InByOut}. These data are processed the same way as other in-transit spectra, and we do not see any signatures of absorption inside the He 10830 \AA~triplet window.

\subsection{Column Density Measurement from Equivalent Width Analysis}
\label{sec:measurement}
We measure the integrated equivalent width ($\mathrm{EW}$) of the detected broad absorption in the vacuum wavelength range (10831.9 \AA\, to 10833.6 \AA\,) in Figure \ref{fig:GJ3470InByOut} to be 0.012 $\pm$0.002\AA\,. 
We ignored the J=0 line (10832.06 \AA\, in vacuum) of the triplet since its oscillator strength is only 1/8$^{th}$ of the other two lines combined, so it is below our continuum noise (J=0 line is the dashed orange line in Figure \ref{fig:GJ3470InByOut}). The resonant scattering absorption of He I 10830 \AA\, is directly proportional to the number density of He I in the $^2_3S$ metastable triplet state. We use the curve of growth analysis in the optically thin regime to estimate the column density of He $^2_3S$ metastable atoms,

\begin{equation}
\label{eq:CurveOfGrowthHe}
N_{He^2_3S}  = \frac{\mathrm{EW}}{8.85 *10^{-13}} \frac{1}{\lambda_2^2 \times f_{ik_2} + \lambda_3^2 \times f_{ik_3}} \unit{cm^{-2}},
\end{equation}
where $N_{He^2_3S}$ is the column density of He $^2_3S$ metastable atoms in $\mathrm{cm^{-2}}$, $\mathrm{EW}$ is the measured combined equivalent width of J=1 and J=2 lines of He 10830 \AA~triplet in units of \AA, $\lambda_2$ and $\lambda_3$ are the wavelengths of the J=1 and J=2 triplet lines in units of microns, and $f_{ik_2}$ and $f_{ik_3}$ are the oscillator strengths of the lines taken from the NIST Atomic Spectra Database Lines Database \citep{drake2006}.

Substituting our measured $\mathrm{EW}$, we obtain $N_{He^2_3S} = 2.4 \times 10^{10} \: \mathrm{cm^{-2}}$. This measured $\mathrm{EW}$ is a flux weighted average across the unresolved stellar disc during the transit. Hence, the column density we measure is also a stellar disc flux weighted average column density of the exosphere during the transit.

\subsection{Theoretical Column Density from an Exosphere Model} 
\label{sec:Tmodel}
We modeled the steady state Helium distribution in the outflowing exosphere of GJ 3470b using the formalism and state transition coefficients outlined in \citet{oklopcic18}. Using hydrodynamic simulations  \citet{salz16} show that the exosphere of GJ 3470b is not isothermal. We therefore do not use a Parker wind model for our atmospheric analysis, but solve for the steady state Helium distribution using the velocity and density field of the exosphere from the \citet{salz16} \texttt{PLUTO-CLOUDY} hydrodynamic simulations. To be internally consistent, we also used the GJ 3470 stellar SED used by \citet{salz16}. 

The X-ray spectrum of GJ 3470 we used from \citet{salz16} is calculated using the plasma emission model from \texttt{CHIANTI} \citep{dere97, dere09} normalised to the X-ray luminosity of GJ 3470b estimated based on the stellar rotation period and the stellar mass \citep{pizzolato03}. The EUV flux (100 \AA~to $\sim$912 \AA), which is critical for these calculations due to the strong dependence of Helium ionization, comes from the scaling of the model-dependent Ly$\alpha$ flux \citep{linsky14}. The model-dependent Ly$\alpha$ flux of GJ 3470 itself is estimated by the \citet{linsky13} model based on the X-ray luminosity. See Section 2.2 of \citet{salz16} for more details. This model-dependent irradiance spectrum of GJ 3470 is the source of the biggest uncertainty in our calculations. However, the model-dependent Ly$\alpha$ flux \citet{salz16} derive, and the value we adopt here, is consistent with the Ly$\alpha$ flux measured by \citet{bourrier18} during three transits of GJ 3470b using the Space Telescope Imaging Spectrograph instrument on board the Hubble Space Telescope (HST). 

As most of the free electrons in the atmospheres of exoplanets come from ionized Hydrogen, to obtain the electron density of GJ 3470 atmosphere, we first solved for the steady-state of Hydrogen following \citet{oklopcic18}. To obtain the steady state distribution of Helium atoms, we briefly discuss here the relevant system of integro-differential equations from \citet{oklopcic18}, given by,
\begin{equation}
\label{eq:HeDistributionOfState1}
\begin{aligned}
v\frac{\partial f_1}{\partial r}  = {} & (1-f_1-f_3)n_e\alpha_1 + f_3 A_{31} - f_1 \Phi_1 e^{-\tau_1} \\
   & - f_1 n_e q_{13a} + f_3 n_e q_{31a} + f_3 n_e q_{31b} \\
   & + f_3 n_{H^0} Q_{31},
 \end{aligned}
\end{equation}
and,
\begin{equation}
\label{eq:HeDistributionOfState3}
\begin{aligned}
v\frac{\partial f_3}{\partial r} = {} & (1-f_1 -f_3)n_e \alpha_3 -f_3 A_{31} - f_3 \Phi_3 e^{-\tau_3} \\
  & + f_1 n_e q_{13a} -f_3 n_e q_{31a} - f_3 n_e q_{31b} \\
  & - f_3 n_{H^o} Q_{31}, 
\end{aligned}
\end{equation}
where $f_1$ and $f_3$ are the fractions of neutral Helium in the ground state of the singlet and triplet states, respectively, $v$ is the velocity of the out-flowing exosphere as a function of radius we adopt from \citet{salz16}, $n_e$ is the density of free electrons we obtained by solving the steady-state of Hydrogen, and $\alpha_1$ and $\alpha_3$ are the Helium recombination rates to singlet and triplet state from \citet{osterbrock06}, respectively. The collision coefficients $q_{ijk}$ and $Q_{31}$ are from \citet{oklopcic18}, calculated using coefficients from \citet{bray00,roberge82}. The radiative decay rate $A_{31}$ from the Helium metastable state to singlet state is adopted from \citet{drake71}. $\Phi_1$ and $\Phi_3$ are the effective photoionization rate coefficients calculated using,
\begin{equation}
\label{eq:PhotoIonisationEqn}
\begin{aligned}
\Phi_i = \int_{\lambda_1}^{\lambda_2} \frac{\lambda F_\lambda}{hc} a_\lambda d\lambda , 
\end{aligned}
\end{equation}
where, $F_\lambda$ is the irradiated flux on GJ3470b described earlier in this section, $a_\lambda$ is the photoionisation cross section taken from \citet{brown71} for the singlet state, and from \citet{norcross71} for the triplet state. For the singlet state, the integral is evaluated up to the ionisation wavelength of Helium (504 \AA), while for the triplet state the integral is computed in the interval starting at Lyman limit\footnote{Photons of energy higher than Lyman limit are many orders of magnitude more likely to be absorbed by the Hydrogen than He$^2_3$S, hence the beginning of the integral window was chosen to be at 911.6 \AA} (911.6 \AA) to the metastable state ionization wavelength (2583 \AA). 

To calculate the optical depths $\tau_1$ and $\tau_3$ as a function of radius for the singlet and metastable states of Helium, we compute the following integrals,
\begin{equation}
\label{eq:OpticalDepthEqn1}
\begin{aligned}
\tau_1 = {} & a_{oH} \frac{0.9}{1.297*m_p} \int_{r}^{\infty} (1-f_{H+}) \rho(r) d r \\
   & + a_{oHe} \frac{0.1}{1.297*m_p} \int_{r}^{\infty} f_1 \rho(r) d r, 
\end{aligned}
\end{equation}
and,
\begin{equation}
\label{eq:OpticalDepthEqn3}
\begin{aligned}
\tau_3 = {} & a_{oHe2S3} \frac{0.1}{1.297*m_p} \int_{r}^{\infty} f_3 \rho(r) d r,
\end{aligned}
\end{equation}
where $m_p$ is the proton mass, $\rho(r)$ is the density of the GJ3470b exosphere from \citet{salz16}, $f_{H+}$ is the fraction of ionized Hydrogen we obtained by solving the steady-state of Hydrogen. $a_{oH}$ and $a_{oHe}$ are GJ3470b's irradiation flux-averaged Hydrogen and Helium ionisation cross sections in the wavelength range up to Helium ionisation (504 \AA). $a_{oHe2S3}$ is the flux-averaged cross section of He $^2_3S$ in the wavelength range 911.6 \AA\, (hydrogen ionisation) to 2583 \AA\, (He $^2_3S$ ionisation).

We solve Equations \ref{eq:HeDistributionOfState1} \& \ref{eq:HeDistributionOfState3} iteratively as an initial boundary value partial differential equation by starting with an initial estimate for the optical depth integral term (Equations \ref{eq:OpticalDepthEqn1} \& \ref{eq:OpticalDepthEqn3}). Using the resulting solution, we updated the integral term in Equations \ref{eq:OpticalDepthEqn1} \& \ref{eq:OpticalDepthEqn3}. The final solutions converged within one or two iterations since the integral term has impact only in the high opacity base region of the outflowing atmosphere. The differential equation of He $^2_3S$ metastable atoms (Equation \ref{eq:HeDistributionOfState3}) is very stiff for high densities of GJ 3470b, and we therefore used a Radau solver with adaptive dense gridding \citep{hairer96}.

\begin{figure*}
  \begin{tabular}[b]{@{}p{0.45\textwidth}@{}}
\begin{center}
\includegraphics[width=0.45\textwidth]{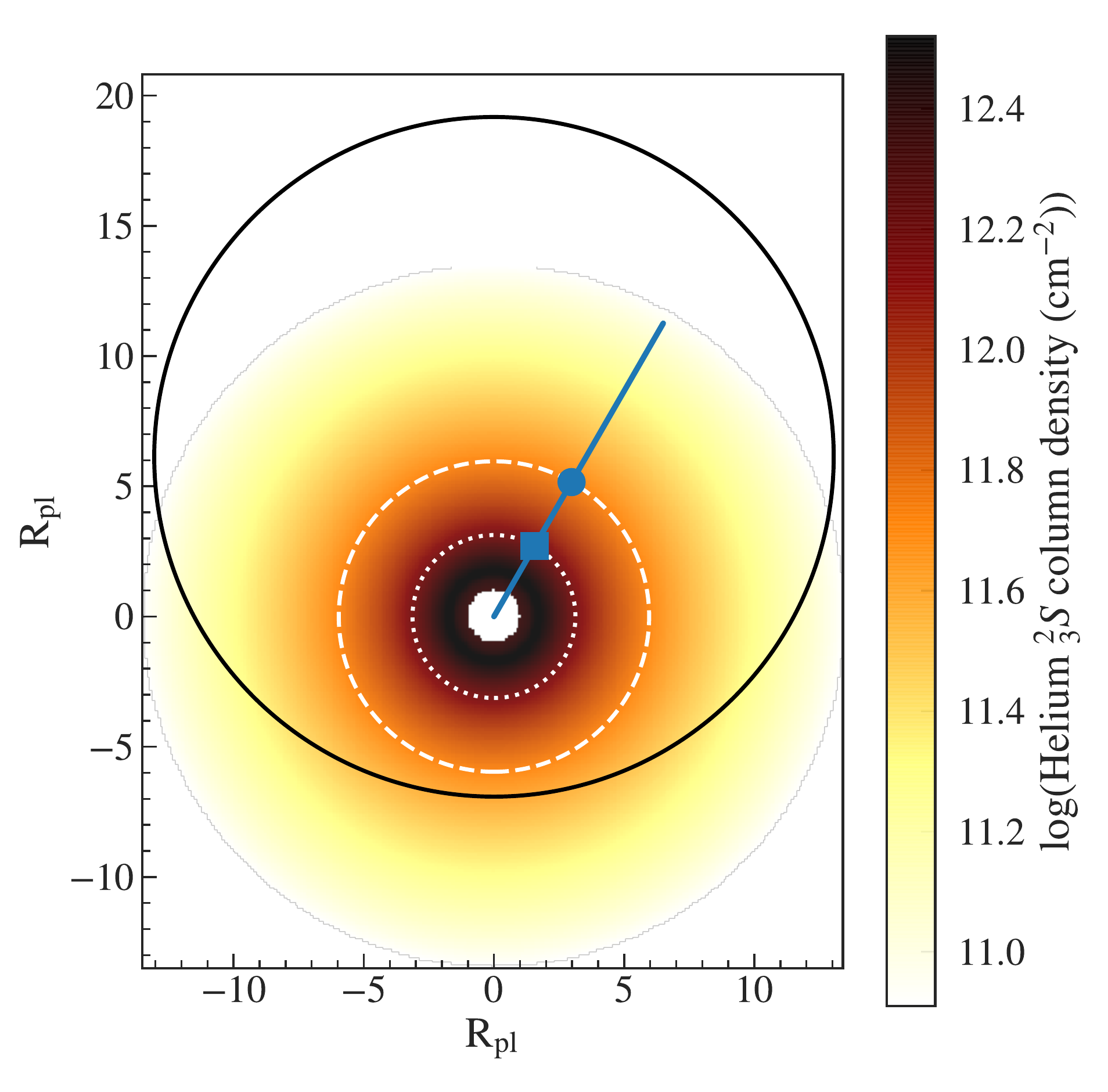}\\ a)
\end{center}
\end{tabular}%
  \quad
  \begin{tabular}[b]{@{}p{0.45\textwidth}@{}}
\begin{center}
\includegraphics[width=0.5\textwidth]{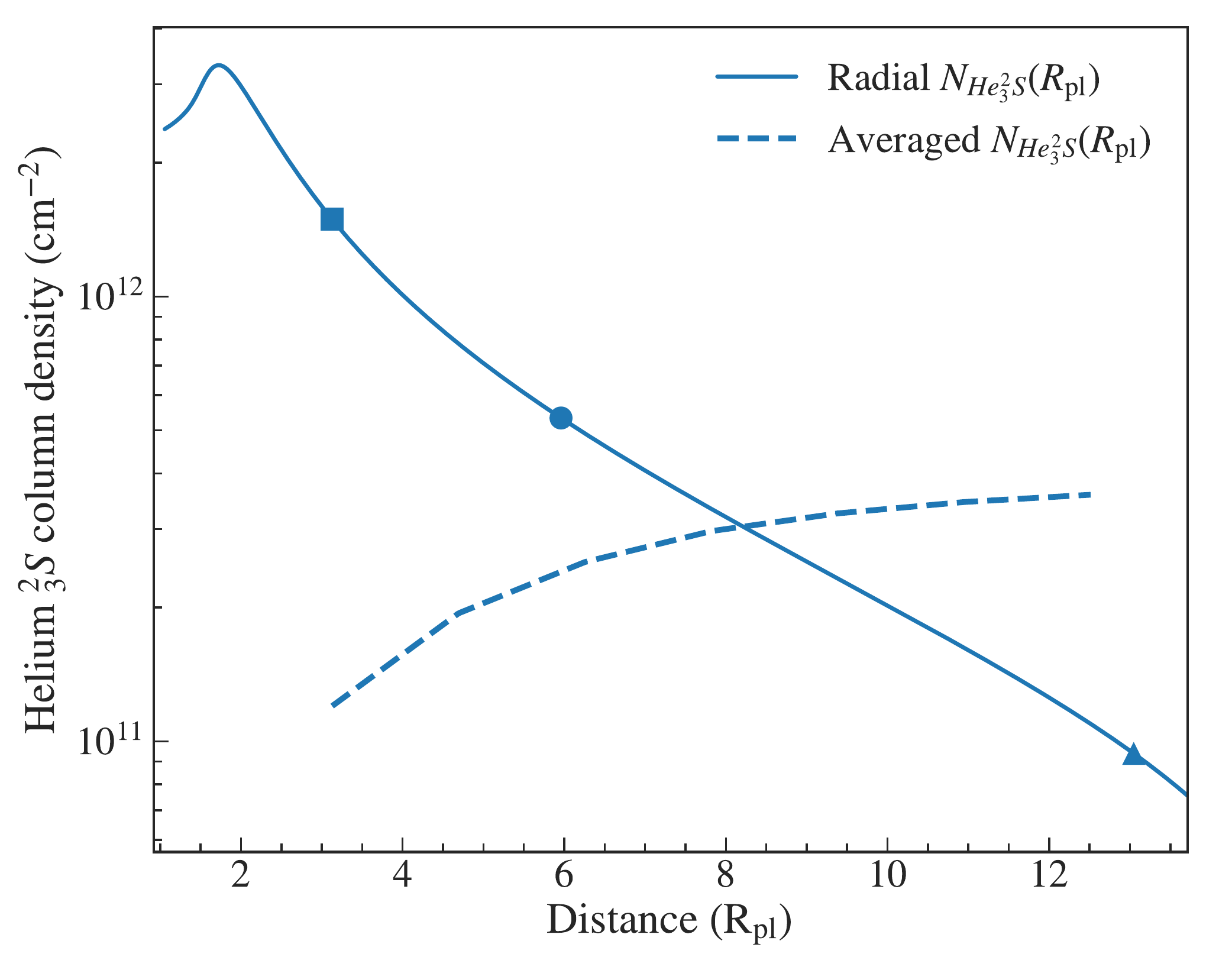}\\ b)
\end{center}
\end{tabular}
\caption{a) The 2D diagram showing the projection of the Roche lobe radii as well as the GJ3470 stellar disc on top of the spherical 1D He $^2_3S$ metastable atom column density map predicted by our 1D simulation of GJ 3470b. The volume equivalent Roche lobe radius as well as the Roche lobe extent on the star-planet axis is marked by the dot and dashed circles respectively. The large circle marks the disc of the star. b) The radial cut plot of the column density along the blue line in left figure is shown here as ``Radial $N_{He^2_3S}$". The volume equivalent Roche lobe radius as well as the Roche lobe extend on the star-planet axis is marked by the square and dot respectively. The triangle marks the radius of the star. The dashed line shows the background stellar disc flux averaged column density predicted by our model at the center of the transit. It is a function of the assumed extent of the exo-sphere beyond the volume-equivalent Roche lobe as shown on left. 
}
\label{fig:He2S3ColumnDensity}
\end{figure*}

The one dimensional radial density distribution of He $^2_3S$ metastable atoms obtained above is then integrated assuming a spherical exosphere to obtain column densities at different impact parameters from the planet (Figure \ref{fig:He2S3ColumnDensity}b). We expect the model to fail beyond the Roche-lobe due to complex stellar wind interactions. We estimate the volume-equivalent Roche-lobe radius of the GJ 3470b system to be 3.12 $R_{p}$ following  \citet{eggleton83}. The teardrop shaped Roche-lobe's extent on the star-planet axis is 5.96 $R_{p}$ \citep{salz16}. Both of these points are explicitly highlighted in Figure \ref{fig:He2S3ColumnDensity}. Assuming an impact parameter of $b=0.47$ \citep{dragomir15}, Figure \ref{fig:He2S3ColumnDensity}a shows the 2D projection of the system during the midpoint of the transit. In Figure \ref{fig:He2S3ColumnDensity}a we also illustrate the stellar disc on top of the column density map of He $^2_3S$ metastable atoms. To estimate the flux-averaged column density, we used a quadratic limb darkening model where we calculated the limb-darkening coefficients using the EXOFAST web-applet\footnote{\url{http://astroutils.astronomy.ohio-state.edu/exofast/limbdark.shtml}} using the stellar effective temperature, metallicity and surface gravity from Table \ref{table:GJ3470properties}, in the $J$ band. This flux-weighted average column density is dependent on how far the exo-sphere extends beyond the Roche-lobe; the predicted background stellar disc flux-weighted average column density of He $^2_3S$ metastable atoms as a function of the extent of GJ 3470b's exosphere is shown by the dashed line in Figure \ref{fig:He2S3ColumnDensity}b).

\subsubsection{Line Profile Model}
\label{sec:LineProfileModel}

\begin{figure*}
  \begin{tabular}[b]{@{}p{0.33\textwidth}@{}}
\begin{center}
\includegraphics[width=0.4\textwidth]{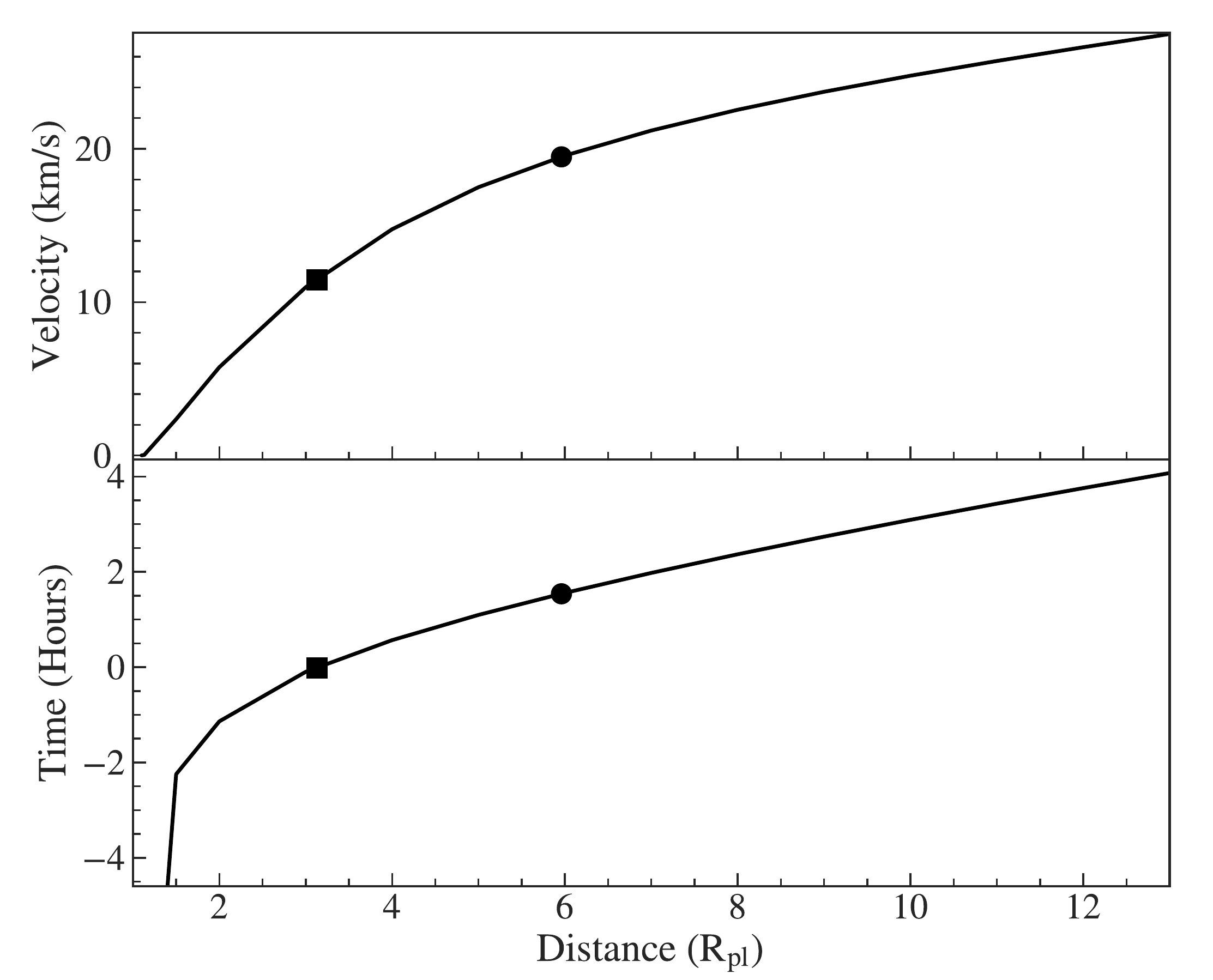}\\ a), b)
\end{center}
\end{tabular}%
  \quad
  \begin{tabular}[b]{@{}p{0.65\textwidth}@{}}
\begin{center}
\includegraphics[width=0.65\textwidth]{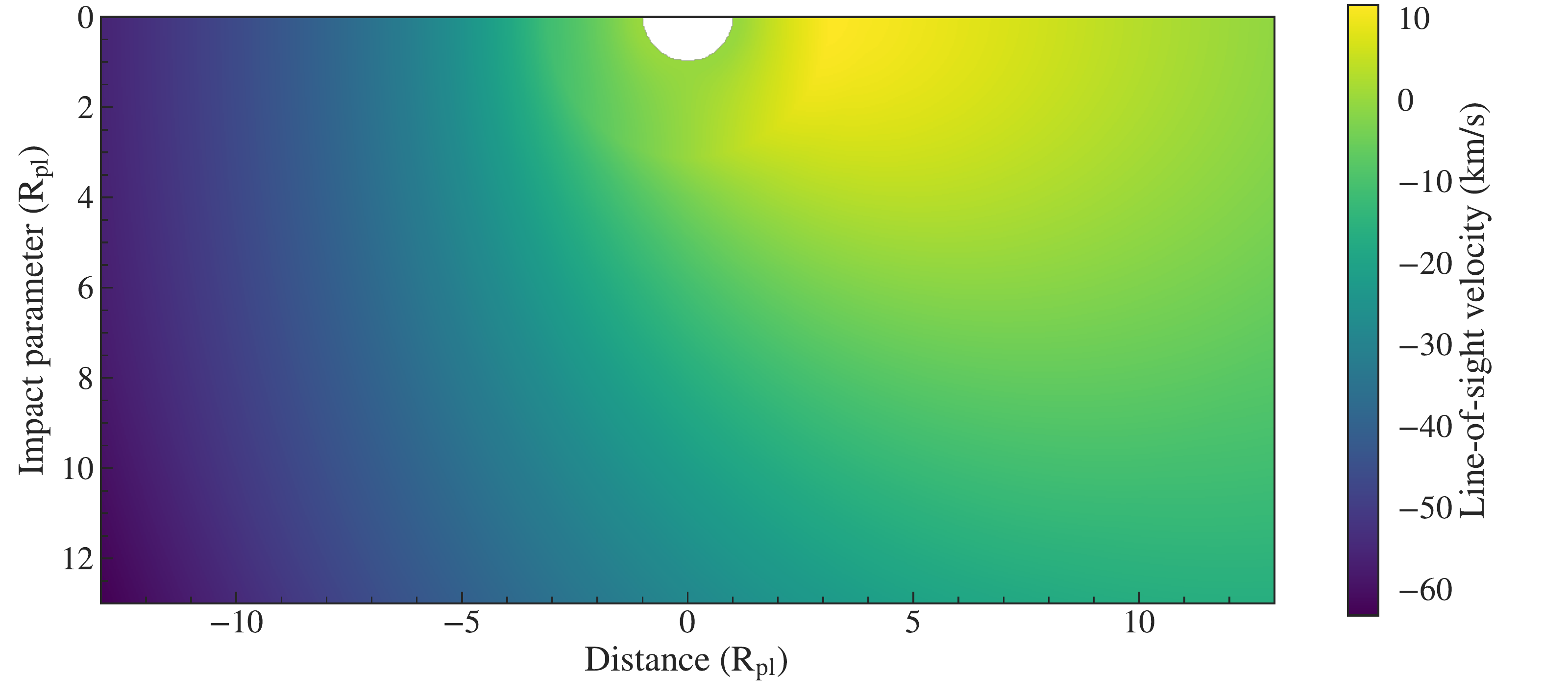}\\ c)
\end{center}
\end{tabular}
\caption{a) The top left panel shows the one dimensional wind velocity model for GJ 3470b from \citet{salz16}. The volume equivalent Roche lobe radius as well as the Roche lobe extent on the star-planet axis is marked by the square and dot respectively. b) The lower left panel shows the time of flight for a Helium atom to reach a certain radius from the volume equivalent Roche lobe radius.  c) The line-of-sight velocity field of exo-sphere around GJ 3470b, Earth observer is towards the left side of the axis and the host star is towards the right side.}
\label{fig:WindVelocitymodel}
\end{figure*}

The broad absorption we present evidence for during the transit of GJ 3470b is possibly  the broadest He 10830 \AA~ absorption reported in the literature to date. 
It spans a velocity range of -36 km/sec to +9 km/sec. Figure \ref{fig:WindVelocitymodel}a shows the outflowing exosphere velocity from \citet{salz16}. The wind reaches only velocities up to $\sim$10 km/s inside 3.12 R$_p$---the volume-equivalent Roche-lobe radius. Both the volume equivalent Roche lobe radius as well as the Roche lobe extent on the star-planet axis (5.96 R$_p$) is marked by vertical lines in the plot.
 
The line of sight velocity of the Helium atoms we see during the transit is the sum of the hydrodynamic driven wind velocity plus the stellar radiative acceleration, and the planet's orbital radial velocity at the instant of their escape from the planet's gravitational potential.  
For example, Helium atoms escaping the planet potential $\sim$2.5 hours before the transit will have a line-of-sight velocity of $\sim$-30 km/s during the transit due to change in the line of sight component of planet's orbital velocity.
Figure \ref{fig:WindVelocitymodel}b shows the time a Helium atom will take to travel the distance based on the wind model after it leaves the Roche lobe. Atoms which escape the potential 2.5 hours before the transit will be at $\sim$8 $R_p$, which is well within the projected stellar disc. 

 For calculating the line profile predicted by our model, we need to calculate the line-of-sight velocity field of the GJ 3470b exo-sphere at all impact parameter positions and radial distances. For the region inside the volume equivalent Roche Lobe, we consider only the line-of-sight projection of the radial hydrodynamic wind velocity. For the region outside the Roche lobe, we added the line-of-sight component of the hydrodynamic wind to the difference in the planet's radial velocity between present and the time at which gas at a certain altitude escaped the Roche lobe. Figure \ref{fig:WindVelocitymodel}c shows this velocity field. The observer on Earth is towards the left side of the axis and the host star is towards the right side. The negative line-of-sight velocity implies gas is moving towards the observer. In reality, there will be additional blue shift due to the interaction with stellar wind. We do not include any stellar wind induced blue shift of the exo-sphere in our line model.

To model the line profile, we used the Voigt profile. The modeling methodology is mostly similar to \citet{oklopcic18}, the only difference is in not assuming a spherically symmetric velocity field, and using the GJ 3470b system's impact parameter and stellar disc limb darkening model while integrating the absorption line profile across the stellar disc. The temperature of the gas for the thermal Gaussian component of the modelled Voigt profile was taken to be 7000 K \citep{salz16,bourrier18}. The He 10830 line's Einstein coefficient for the Lorentzian component of the Voigt profile was taken from NIST Atomic Spectra Database Lines Database \citep{drake2006}. The oscillator strengths for the cross section calculation of each triplet line is also taken from the NIST Database. These cross sections, along with the radial density distribution of the meta stable Helium we obtained in our model (Section \ref{sec:Tmodel}) were integrated to obtain the optical depth at different impact parameter distances from the planet. Density of meta-stable Helium outside 13 R$_{pl}$ (radial limit of our 1D model) was set to zero in our integrals. Using a quadratic limb darkening model, we averaged the transmission curve across the disc to obtain the net transmission profile of the He 10830 triplet lines in GJ 3470b during the transit. Since the measured column density is less than the model predicted column density by a factor of 10 (see Section \ref{sec:eqwcomparison}), we scaled down the density distribution from our model also by a factor of 10. 
This predicted line profile is over plotted over all the transit and out of transit ratio spectra in Figure \ref{fig:GJ3470InByOut}.

\subsubsection{Limits of the Model}
\label{sec:caveats}
The simplified one dimensional model we presented here is to check whether the strength of our claimed Helium 10830 \AA\, absorption in the exo-planet atmosphere is within the limits of expected physical conditions.
A true three dimensional wind model where the stellar irradence is only on one side of planet is known to have lesser density on an average in all angles by a factor of 4 than the one dimensional symmetric wind model we used \citep{stone09}.  
 
\begin{figure}
\begin{center}
\includegraphics[width=1.0\columnwidth]{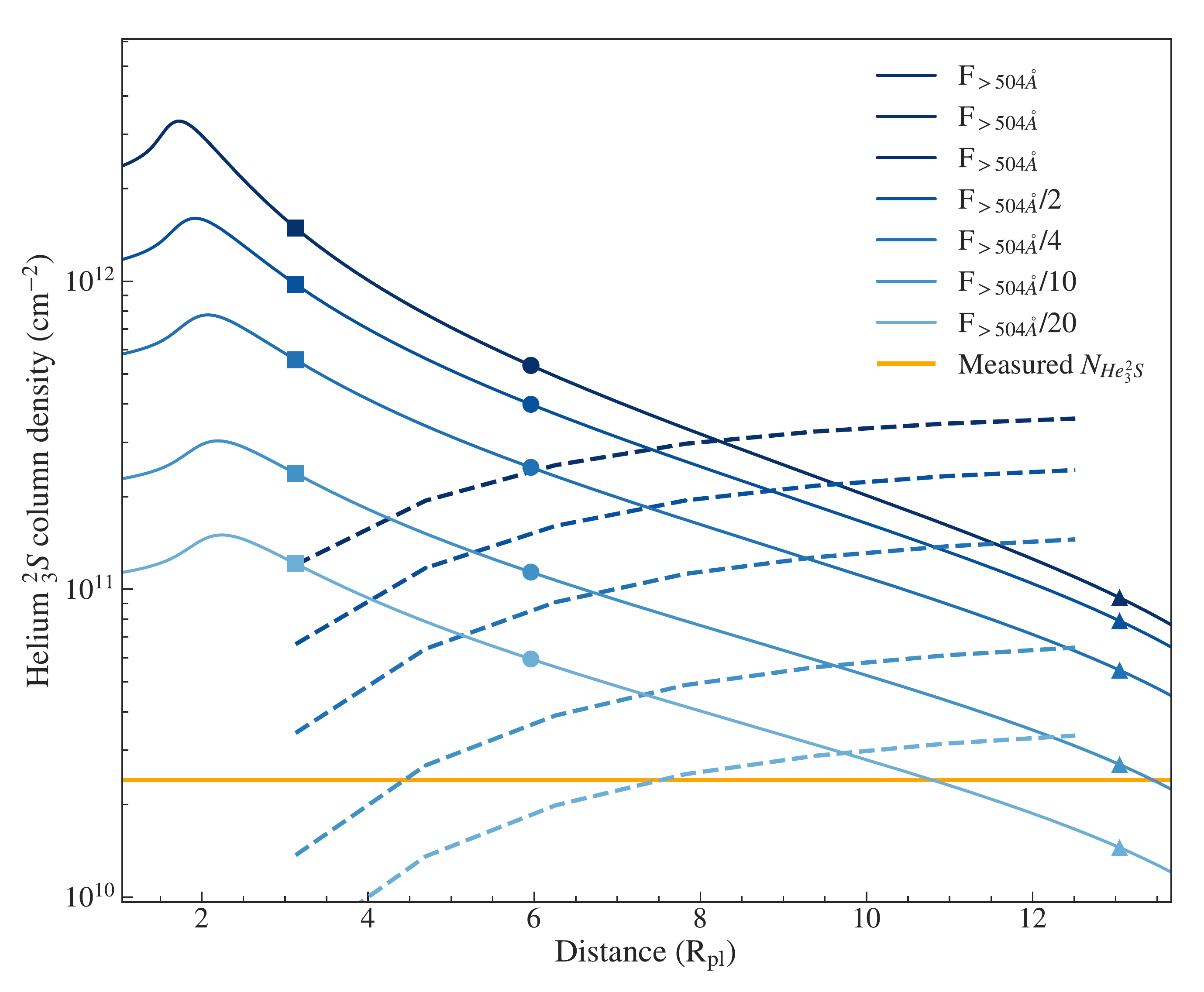}
\caption{The column density of He $^2_3S$ metastable atoms predicted by the 1D simulation of GJ 3470b is reduced proportionally to the flux in wavelengths less than 504 \AA~when everything else---including the density and velocity---is kept fixed. The volume equivalent Roche lobe radius as well as the Roche lobe extent on the star-planet axis is marked by the square and dot respectively. The triangle marks the radius of the star. The dashed lines show the flux averaged column density predicted by this model at the center of the transit as a function of the extent of the exo-sphere beyond volume-equivalent Roche lobe. The horizontal orange line shows the measured average column density in transits of GJ 3470b using HPF.}
\label{fig:FluxDependenceofHe}
\end{center}
\end{figure}

The other major source of uncertainty in our model is the irradiated flux in the wavelengths shorter than Helium ionization ($\lambda < 512$ \AA). As briefly outlined in the section above, it is estimated by a chain of empirical models. We expect the Helium ionization flux to have a significant influence on the population of meta-stable He atoms. To study the impact, we simulated how the predicted column density of He $^2_3S$ metastable atoms changes when we suppress the irradience at wavelength shorter than 504 \AA~by different factors\footnote{The reason we chose 504 \AA\, for this test instead of 512 \AA\, is because photons of energy higher than 504 \AA\, are more likely to ionize a Helium atom than Hydrogen atom due to the factor of 10 lower abundance of Helium than Hydrogen. We want to probe the impact on Helium ionization alone without significant change to Hydrogen ionization.}. Results are shown in Figure \ref{fig:FluxDependenceofHe}. Since we wanted to probe only the impact on the steady state of Helium atoms, the underlying velocity field and density used in the differential equation was kept fixed, and is the same as the \citet{salz16} hydrodynamic model for the original irradiance. It is very instructive to see that the observed column density of metastable Helium is reduced proportionally to the reduction in the Helium ionization flux. 

In real systems, since the mass outflow is proportional to the energy absorbed in the lower atmosphere from irradiance, there will be an additional proportional reduction in density when the EUV flux irradiation is reduced. Hence, the combined effect of reduction in Helium ionizing radiation is quadratic on the column density of meta-stable Helium atoms.

A few more ways the underlying density of the wind we used in our simulation from \citet{salz16} can be impacted is summarized in Table 2 of \citet{salz16}.

\subsection{On the Difference Between the Predicted and Measured Column Densities}
\label{sec:eqwcomparison}
As described in Section \ref{sec:measurement}, the measured flux weighted column density of He $^2_3$S is $N_{He^2_3S} = 2.4 \times 10^{10} \mathrm{cm^{-2}}$. 
This is a factor of 10 less than predicted by our one dimensional hydrodynamic model described in Section \ref{sec:Tmodel}. 
However, as discussed in Section \ref{sec:caveats}, if we correct our simulation for the 3D model simplification to 1D model by the factor of 4 \citep{stone09},  and assume the Helium ionisation radiation is lesser by a factor of 1.6 (which is well within the error of the empirically derived EUV SED of GJ3470 in \citet{salz16}), we reduce the predicted metastable Helium by a factor of $4\times1.6^2 = 10$. Hence, our observations are consistent with our simplified model under the caveats outlined in Section \ref{sec:caveats}.

\subsection{On the Difference in Model Line Profile and the Observed Absorption Signature}
\label{sec:lineprofileshapedifference}
 We discuss a few possible reasons for differences in the shape of the simple model and observed line profile below.

{{\bf Stellar Activity:} 
Some M dwarfs can be active, and exhibit time variability in He 10830 absorption line (we note that not all M dwarfs are active). In comparison to other transmission spectroscopy lines like Ca II K, Na I D and H$\alpha$, active regions typically have He 10830 in absorption \citep{cauley18}. Thus, when a planet transits an active spot region on the host star, He 10830 is more likely to result in an emission signature, than an absorption signature. Thus He 10830 absorption signatures in transit spectroscopy are more likely to originate in the exo-sphere than from the stellar chromosphere.  However, under high density chromosphere conditions \citep{andretta95}, or active flares, it is possible to obtain He 10830 in emission from the recombination cascade of ionised Helium \citep{huenemoerder87}. We do not see any indication of flares during transit in the Ca II IR triplet lines. If the absorption we detected was due to GJ 3470b transiting a very active emission spot, we would expect the absorption signal to not be present in all the in transit spectra taken (see Figure \ref{fig:IndividualRatioSpectrum}). We have a total of 9 in transit spectra, (3 spectra from 3 transits each). Since the phase of each spectrum is slightly different, it is improbable that there was an unusually strong emission spot of He 10830 during each of our transits. Hence, stellar activity cannot explain the absorption signal we detected during transit of GJ3470b. Stellar activity, however, could probably explain some of the emission structures, and there by the net absorption strength variations we measured across the three transits.

\begin{figure}
\begin{center}
\includegraphics[width=1.0\columnwidth]{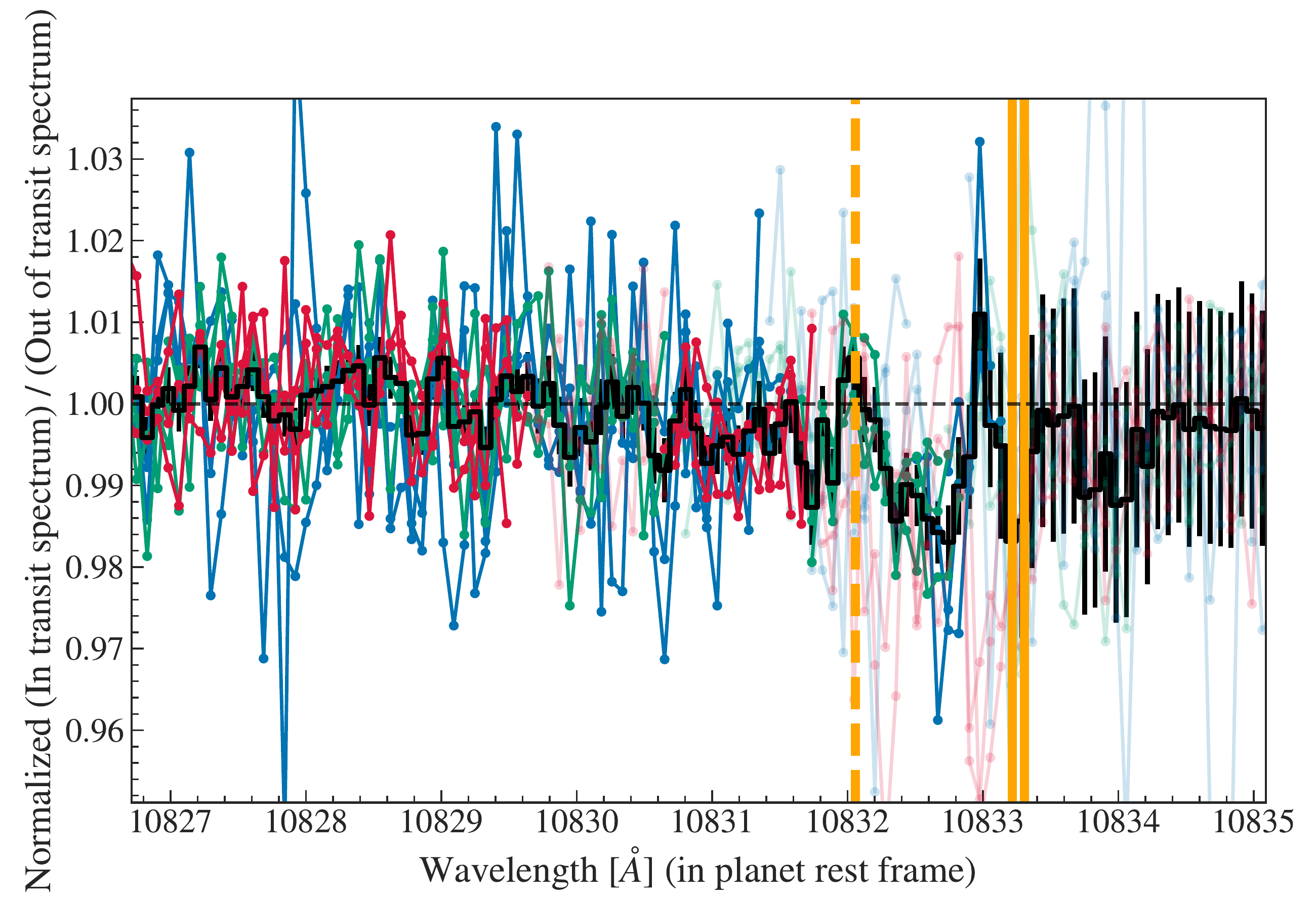}
\caption{ Individual in-transit by out-of-transit ratio HPF spectrum of all 9 epochs from the three transits are shown here. Blue curves are the three spectra from the first transit, green are the second transit spectra, and the red are the third transit spectra. The black curve is the weighted average with error bars of all the in-transit by out-of-transit spectra. Translucent parts of the curves are telluric and sky emission line corrected regions in each spectrum. They are de-weighted to remove any systematic artifacts from influencing the final average. The x-axis shows vacuum wavelength in the planet’s rest frame at mid-transit. The rest vacuum wavelengths of the He 10830 \AA\, triplet lines in planet's rest frame are marked by the vertical dash and solid orange lines.}
\label{fig:IndividualRatioSpectrum}
\end{center}
\end{figure}

{\bf Telluric Absorption \& Sky Emission:} Systematics in telluric and sky emission line modelling could potentially cause profile variations in the regions with high error bars in Figure \ref{fig:GJ3470InByOut}. The regions with small error bars are free of any telluric or sky emission line contamination in at least one epoch out of the three transit observations. Hence, imperfect telluric line modelling errors cannot explain the absorption signature.

{\bf Stellar Wind:} Even though our simple model which includes the orbital dynamics and the hydro-dynamic outflow velocity could explain the large blue shifted profile seen in the data, it does not fully explain the shape of the detected line. Some component of this discrepancy could be variability, however, we suspect the stellar wind to play a major role. Since stellar wind accelerates the exo-sphere towards earth like a comet tail, it could further blue shift the exo-sphere at the base of the exo-sphere skewing the line profile to shorter wavelengths. Further monitoring of the GJ 3470b's He 10830 signature during transits will reveal greater insights on the system.}

\section{Conclusion}
\label{sec:concl}
We report evidence for He 10830 \AA~absorption from meta-stable Helium atoms in the base of the outflowing exosphere of the M dwarf planet GJ 3470b in three transits using the Habitable-zone Planet Finder (HPF) near-infrared spectrograph at the 10m Hobby-Eberly Telescope at McDonald Observatory. This measurement marks the first evidence for He 10830 absorption in an M-dwarf planet. Further, we detect the Helium absorption in the velocity range of -36 km/sec to +9 km/sec, marking the largest blueshift of He 10830 absorption reported so far in the literature. From our observed absorption, we measure an equivalent width of $\mathrm{EW} = 0.012 \pm 0.002 \AA$, corresponding to a disc surface flux-averaged column density of $N_{He^2_3S} = 2.4 \times 10^{10} \unit{cm^{-2}}$. Both the velocity range and the column density we measure are consistent with our exosphere model based on the work of \citet{salz16} and \citet{oklopcic18} to within the model uncertainties of the UV and X-ray flux of GJ 3470. 
 These observations suggest that detection of the He 10830 absorption around hot Neptunes is now within reach of ground based facilities - opening up a new window to probe these systems. Puzzling aspects of the broad absorption seen remains- which we have attempted to discuss the sources of.  We recommend future ground and space based observations to probe the physics of this absorption signal at even higher signal to noise.

\acknowledgements
The authors wish to thank M. Salz and A. Oklop{\v{c}}i{\'c} for discussions that contributed to this work. We thank the anonymous referee for a thoughtful reading of the manuscript and their suggestions that improved the quality of the manuscript.\\
This work was partially supported by funding from the Center for Exoplanets and Habitable Worlds. The Center for Exoplanets and Habitable Worlds is supported by the Pennsylvania State University, the Eberly College of Science, and the Pennsylvania Space Grant Consortium.  We acknowledge support from NSF grants AST-1006676, AST-1126413, AST-1310885, AST-1517592, AST-1310875, the NASA Astrobiology Institute (NAI; NNA09DA76A), PSARC, and NIST in our pursuit of precision radial velocities in the NIR. Computations for this research were performed on the Pennsylvania State University’s Institute for CyberScience Advanced CyberInfrastructure (ICS-ACI). We acknowledge support from the Heising-Simons Foundation via grant 2017-0494. This work was supported by NASA Headquarters under the NASA Earth and Space Science Fellowship Program through grants NNX16AO28H and 80NSSC18K1114.

These results are based on observations obtained with the Habitable-zone Planet Finder Spectrograph on the Hobby-Eberly Telescope. We thank the Resident astronomers and Telescope Operators at the HET for the skillful execution of our observations of our observations with HPF. The Hobby-Eberly Telescope is a joint project of the University of Texas at Austin, the Pennsylvania State University, Ludwig-Maximilians-Universität München, and Georg-August Universität Gottingen. The HET is named in honor of its principal benefactors, William P. Hobby and Robert E. Eberly. The HET collaboration acknowledges the support and resources from the Texas Advanced Computing Center.

This research has made use of NASA's Astrophysics Data System Bibliographic Services.

Software: barycorrpy \citep{kanodia18}, astropy \citep{Astropy2018}, numpy \citep{Numpy2011}, scipy \citep{Scipy2001}, matplotlib \citep{Hunter2007}, CoCalc \citep{sage}, GNU parallel \citep{Tange2011a}.

\facilities{HET (HPF)}

\bibliography{HPFreferences}

\begin{thebibliography}{}
\expandafter\ifx\csname natexlab\endcsname\relax\def\natexlab#1{#1}\fi
\providecommand{\url}[1]{\href{#1}{#1}}

\bibitem[{{Allart} {et~al.}(2018){Allart}, {Bourrier}, {Lovis}, {Ehrenreich},
  {Spake}, {Wyttenbach}, {Pino}, {Pepe}, {Sing}, \& {Lecavelier des
  Etangs}}]{allart18}
{Allart}, R., {Bourrier}, V., {Lovis}, C., {et~al.} 2018, Science, 362, 1384

\bibitem[{{Allart} {et~al.}(2019){Allart}, {Bourrier}, {Lovis}, {Ehrenreich},
  {Aceituno}, {Guijarro}, {Pepe}, {Sing}, {Spake}, \& {Wyttenbach}}]{allart19}
---. 2019, \aap, 623, A58

\bibitem[{{Alonso-Floriano} {et~al.}(2019){Alonso-Floriano}, {Snellen},
  {Czesla}, {Bauer}, {Salz}, {Lamp{\'o}n}, {Lara}, {Nagel},
  {L{\'o}pez-Puertas}, {Nortmann}, {S{\'a}nchez-L{\'o}pez}, {Sanz-Forcada},
  {Caballero}, {Reiners}, {Ribas}, {Quirrenbach}, {Amado}, {Aceituno},
  {Anglada-Escud{\'e}}, {B{\'e}jar}, {Brinkm{\"o}ller}, {Hatzes}, {Henning},
  {Kaminski}, {K{\"u}rster}, {Labarga}, {Montes}, {Pall{\'e}}, {Schmitt}, \&
  {Zapatero Osorio}}]{floriano19}
{Alonso-Floriano}, F.~J., {Snellen}, I.~A.~G., {Czesla}, S., {et~al.} 2019,
  \aap, 629, A110

\bibitem[{{Andretta} \& {Giampapa}(1995)}]{andretta95}
{Andretta}, V., \& {Giampapa}, M.~S. 1995, \apj, 439, 405

\bibitem[{{Astropy Collaboration} {et~al.}(2018){Astropy Collaboration},
  {Price-Whelan}, {Sip{\H o}cz}, {G{\"u}nther}, {Lim}, {Crawford}, {Conseil},
  {Shupe}, {Craig}, {Dencheva}, {Ginsburg}, {VanderPlas}, {Bradley},
  {P{\'e}rez-Su{\'a}rez}, {de Val-Borro}, {Aldcroft}, {Cruz}, {Robitaille},
  {Tollerud}, {Ardelean}, {Babej}, {Bach}, {Bachetti}, {Bakanov}, {Bamford},
  {Barentsen}, {Barmby}, {Baumbach}, {Berry}, {Biscani}, {Boquien}, {Bostroem},
  {Bouma}, {Brammer}, {Bray}, {Breytenbach}, {Buddelmeijer}, {Burke},
  {Calderone}, {Cano Rodr{\'{\i}}guez}, {Cara}, {Cardoso}, {Cheedella},
  {Copin}, {Corrales}, {Crichton}, {D'Avella}, {Deil}, {Depagne}, {Dietrich},
  {Donath}, {Droettboom}, {Earl}, {Erben}, {Fabbro}, {Ferreira}, {Finethy},
  {Fox}, {Garrison}, {Gibbons}, {Goldstein}, {Gommers}, {Greco}, {Greenfield},
  {Groener}, {Grollier}, {Hagen}, {Hirst}, {Homeier}, {Horton}, {Hosseinzadeh},
  {Hu}, {Hunkeler}, {Ivezi{\'c}}, {Jain}, {Jenness}, {Kanarek}, {Kendrew},
  {Kern}, {Kerzendorf}, {Khvalko}, {King}, {Kirkby}, {Kulkarni}, {Kumar},
  {Lee}, {Lenz}, {Littlefair}, {Ma}, {Macleod}, {Mastropietro}, {McCully},
  {Montagnac}, {Morris}, {Mueller}, {Mumford}, {Muna}, {Murphy}, {Nelson},
  {Nguyen}, {Ninan}, {N{\"o}the}, {Ogaz}, {Oh}, {Parejko}, {Parley}, {Pascual},
  {Patil}, {Patil}, {Plunkett}, {Prochaska}, {Rastogi}, {Reddy Janga},
  {Sabater}, {Sakurikar}, {Seifert}, {Sherbert}, {Sherwood-Taylor}, {Shih},
  {Sick}, {Silbiger}, {Singanamalla}, {Singer}, {Sladen}, {Sooley},
  {Sornarajah}, {Streicher}, {Teuben}, {Thomas}, {Tremblay}, {Turner},
  {Terr{\'o}n}, {van Kerkwijk}, {de la Vega}, {Watkins}, {Weaver}, {Whitmore},
  {Woillez}, {Zabalza}, \& {Astropy Contributors}}]{Astropy2018}
{Astropy Collaboration}, {Price-Whelan}, A.~M., {Sip{\H o}cz}, B.~M., {et~al.}
  2018, \aj, 156, 123

\bibitem[{{Awiphan} {et~al.}(2016){Awiphan}, {Kerins}, {Pichadee},
  {Komonjinda}, {Dhillon}, {Rujopakarn}, {Poshyachinda}, {Marsh}, {Reichart},
  {Ivarsen}, \& {Haislip}}]{awiphan16}
{Awiphan}, S., {Kerins}, E., {Pichadee}, S., {et~al.} 2016, \mnras, 463, 2574

\bibitem[{{Bender} {et~al.}(2012){Bender}, {Mahadevan}, {Deshpande}, {Wright},
  {Roy}, {Terrien}, {Sigurdsson}, {Ramsey}, {Schneider}, \&
  {Fleming}}]{bender2012}
{Bender}, C.~F., {Mahadevan}, S., {Deshpande}, R., {et~al.} 2012, \apjl, 751,
  L31

\bibitem[{{Biddle} {et~al.}(2014){Biddle}, {Pearson}, {Crossfield}, {Fulton},
  {Ciceri}, {Eastman}, {Barman}, {Mann}, {Henry}, {Howard}, {Williamson},
  {Sinukoff}, {Dragomir}, {Vican}, {Mancini}, {Southworth}, {Greenberg},
  {Turner}, {Thompson}, {Taylor}, {Levine}, \& {Webber}}]{biddle14}
{Biddle}, L.~I., {Pearson}, K.~A., {Crossfield}, I.~J.~M., {et~al.} 2014,
  \mnras, 443, 1810

\bibitem[{{Bourrier} {et~al.}(2018){Bourrier}, {Lecavelier des Etangs},
  {Ehrenreich}, {Sanz-Forcada}, {Allart}, {Ballester}, {Buchhave}, {Cohen},
  {Deming}, {Evans}, {Garc{\'\i}a Mu{\~n}oz}, {Henry}, {Kataria}, {Lavvas},
  {Lewis}, {L{\'o}pez-Morales}, {Marley}, {Sing}, \& {Wakeford}}]{bourrier18}
{Bourrier}, V., {Lecavelier des Etangs}, A., {Ehrenreich}, D., {et~al.} 2018,
  \aap, 620, A147

\bibitem[{{Bray, I.} {et~al.}(2000){Bray, I.}, {Burgess, A.}, {Fursa, D. V.},
  \& {Tully, J. A.}}]{bray00}
{Bray, I.}, {Burgess, A.}, {Fursa, D. V.}, \& {Tully, J. A.} 2000, Astron.
  Astrophys. Suppl. Ser., 146, 481.
\newblock \url{https://doi.org/10.1051/aas:2000277}

\bibitem[{{Brown}(1971)}]{brown71}
{Brown}, R.~L. 1971, \apj, 164, 387

\bibitem[{{Cauley} {et~al.}(2018){Cauley}, {Kuckein}, {Redfield}, {Shkolnik},
  {Denker}, {Llama}, \& {Verma}}]{cauley18}
{Cauley}, P.~W., {Kuckein}, C., {Redfield}, S., {et~al.} 2018, \aj, 156, 189

\bibitem[{Clough {et~al.}(2005)Clough, Shephard, Mlawer, Delamere, Iacono,
  Cady-Pereira, Boukabara, \& Brown}]{clough05}
Clough, S., Shephard, M., Mlawer, E., {et~al.} 2005, Journal of Quantitative
  Spectroscopy and Radiative Transfer, 91, 233 .
\newblock
  \url{http://www.sciencedirect.com/science/article/pii/S0022407304002158}

\bibitem[{{Dere} {et~al.}(1997){Dere}, {Landi}, {Mason}, {Monsignori Fossi}, \&
  {Young}}]{dere97}
{Dere}, K.~P., {Landi}, E., {Mason}, H.~E., {Monsignori Fossi}, B.~C., \&
  {Young}, P.~R. 1997, \aaps, 125, 149

\bibitem[{{Dere} {et~al.}(2009){Dere}, {Landi}, {Young}, {Del Zanna},
  {Landini}, \& {Mason}}]{dere09}
{Dere}, K.~P., {Landi}, E., {Young}, P.~R., {et~al.} 2009, \aap, 498, 915

\bibitem[{{Dragomir} {et~al.}(2015){Dragomir}, {Benneke}, {Pearson},
  {Crossfield}, {Eastman}, {Barman}, \& {Biddle}}]{dragomir15}
{Dragomir}, D., {Benneke}, B., {Pearson}, K.~A., {et~al.} 2015, \apj, 814, 102

\bibitem[{Drake(2006)}]{drake2006}
Drake, G. 2006, High Precision Calculations for Helium, ed. G.~Drake (New York,
  NY: Springer New York), 199--219.
\newblock \url{https://doi.org/10.1007/978-0-387-26308-3_11}

\bibitem[{Drake(1971)}]{drake71}
Drake, G. W.~F. 1971, Phys. Rev. A, 3, 908.
\newblock \url{https://link.aps.org/doi/10.1103/PhysRevA.3.908}

\bibitem[{{Eggleton}(1983)}]{eggleton83}
{Eggleton}, P.~P. 1983, \apj, 268, 368

\bibitem[{Ehrenreich {et~al.}(2015)Ehrenreich, Bourrier, Wheatley, des Etangs,
  Hébrard, Udry, Bonfils, Delfosse, Désert, Sing, \&
  Vidal-Madjar}]{ehrenreich15}
Ehrenreich, D., Bourrier, V., Wheatley, P.~J., {et~al.} 2015, Nature, 522, 459.
\newblock \url{https://www.nature.com/articles/nature14501}

\bibitem[{{Gaia Collaboration} {et~al.}(2018){Gaia Collaboration}, {Brown},
  {Vallenari}, {Prusti}, {de Bruijne}, {Babusiaux}, {Bailer-Jones}, {Biermann},
  {Evans}, {Eyer}, \& et~al.}]{gaiaDR218}
{Gaia Collaboration}, {Brown}, A.~G.~A., {Vallenari}, A., {et~al.} 2018, \aap,
  616, A1

\bibitem[{Hairer \& Wanner(1996)}]{hairer96}
Hairer, E., \& Wanner, G. 1996, Solving Ordinary Differential Equations II.
  Stiff and Differential-Algebraic Problems, Vol.~14,
  doi:10.1007/978-3-662-09947-6

\bibitem[{{Huenemoerder} \& {Ramsey}(1987)}]{huenemoerder87}
{Huenemoerder}, D.~P., \& {Ramsey}, L.~W. 1987, \apj, 319, 392

\bibitem[{{Hunter}(2007)}]{Hunter2007}
{Hunter}, J.~D. 2007, Computing in Science and Engineering, 9, 90

\bibitem[{Jones {et~al.}(2001--)Jones, Oliphant, Peterson,
  {et~al.}}]{Scipy2001}
Jones, E., Oliphant, T., Peterson, P., {et~al.} 2001--, {SciPy}: Open source
  scientific tools for {Python}, , .
\newblock \url{http://www.scipy.org/}

\bibitem[{{Kanodia} \& {Wright}(2018)}]{kanodia18}
{Kanodia}, S., \& {Wright}, J. 2018, Research Notes of the American
  Astronomical Society, 2, 4

\bibitem[{{Kaplan} {et~al.}(2018){Kaplan}, {Bender}, {Terrien}, {Ninan}, {Roy},
  \& {Mahadevan}}]{kaplan2018}
{Kaplan}, K.~F., {Bender}, C.~F., {Terrien}, R., {et~al.} 2018, in The 28th
  International Astronomical Data Analysis Software \& Systems

\bibitem[{{Kosiarek} {et~al.}(2019){Kosiarek}, {Crossfield},
  {Hardegree-Ullman}, {Livingston}, {Benneke}, {Henry}, {Howard}, {Berardo},
  {Blunt}, {Fulton}, {Hirsch}, {Howard}, {Isaacson}, {Petigura}, {Sinukoff},
  {Weiss}, {Bonfils}, {Dressing}, {Knutson}, {Schlieder}, {Werner}, {Gorjian},
  {Krick}, {Morales}, {Astudillo-Defru}, {Almenara}, {Delfosse}, {Forveille},
  {Lovis}, {Mayor}, {Murgas}, {Pepe}, {Santos}, {Udry}, {Corbett}, {Fors},
  {Law}, {Ratzloff}, \& {del Ser}}]{kosiarek19}
{Kosiarek}, M.~R., {Crossfield}, I.~J.~M., {Hardegree-Ullman}, K.~K., {et~al.}
  2019, \aj, 157, 97

\bibitem[{{Kulow} {et~al.}(2014){Kulow}, {France}, {Linsky}, \&
  {Loyd}}]{kulow14}
{Kulow}, J.~R., {France}, K., {Linsky}, J., \& {Loyd}, R.~O.~P. 2014, \apj,
  786, 132

\bibitem[{{Linsky} {et~al.}(2014){Linsky}, {Fontenla}, \& {France}}]{linsky14}
{Linsky}, J.~L., {Fontenla}, J., \& {France}, K. 2014, \apj, 780, 61

\bibitem[{{Linsky} {et~al.}(2013){Linsky}, {France}, \& {Ayres}}]{linsky13}
{Linsky}, J.~L., {France}, K., \& {Ayres}, T. 2013, \apj, 766, 69

\bibitem[{{Mahadevan} {et~al.}(2014){Mahadevan}, {Ramsey}, {Terrien},
  {Halverson}, {Roy}, {Hearty}, {Levi}, {Stefansson}, {Robertson}, {Bender},
  {Schwab}, \& {Nelson}}]{mah2014SPIE}
{Mahadevan}, S., {Ramsey}, L.~W., {Terrien}, R., {et~al.} 2014, in \procspie,
  Vol. 9147, Ground-based and Airborne Instrumentation for Astronomy V, 91471G

\bibitem[{{Mansfield} {et~al.}(2018){Mansfield}, {Bean}, {Oklop{\v c}i{\'c}},
  {Kreidberg}, {D{\'e}sert}, {Kempton}, {Line}, {Fortney}, {Henry}, {Mallonn},
  {Stevenson}, {Dragomir}, {Allart}, \& {Bourrier}}]{mansfield18}
{Mansfield}, M., {Bean}, J.~L., {Oklop{\v c}i{\'c}}, A., {et~al.} 2018, \apjl,
  868, L34

\bibitem[{Metcalf {et~al.}(2019)Metcalf, Anderson, Bender, Blakeslee, Brand,
  Carlson, Cochran, Diddams, Endl, Fredrick, Halverson, Hickstein, Hearty,
  Jennings, Kanodia, Kaplan, Levi, Lubar, Mahadevan, Monson, Ninan, Nitroy,
  Osterman, Papp, Quinlan, Ramsey, Robertson, Roy, Schwab, Sigurdsson,
  Srinivasan, Stefansson, Sterner, Terrien, Wolszczan, Wright, \&
  Ycas}]{metcalf19}
Metcalf, A.~J., Anderson, T., Bender, C.~F., {et~al.} 2019, Optica, 6, 233.
\newblock
  \url{http://www.osapublishing.org/optica/abstract.cfm?URI=optica-6-2-233}

\bibitem[{Moutou {et~al.}(2003)Moutou, Coustenis, Schneider, Queloz, \&
  Mayor}]{moutou03}
Moutou, C., Coustenis, A., Schneider, J., Queloz, D., \& Mayor, M. 2003,
  Astronomy \& Astrophysics, 405, 341.
\newblock
  \url{https://www.aanda.org/articles/aa/abs/2003/25/aa3416/aa3416.html}

\bibitem[{{Ninan} {et~al.}(2018){Ninan}, {Bender}, {Mahadevan}, {Ford},
  {Monson}, {Kaplan}, {Terrien}, {Roy}, {Robertson}, {Kanodia}, \&
  {Stefansson}}]{ninan18}
{Ninan}, J.~P., {Bender}, C.~F., {Mahadevan}, S., {et~al.} 2018, in Society of
  Photo-Optical Instrumentation Engineers (SPIE) Conference Series, Vol. 10709,
  High Energy, Optical, and Infrared Detectors for Astronomy VIII, 107092U

\bibitem[{Norcross(1971)}]{norcross71}
Norcross, D.~W. 1971, Journal of Physics B: Atomic and Molecular Physics, 4,
  652.
\newblock \url{https://doi.org/10.1088%2F0022-3700%2F4%2F5%2F006}

\bibitem[{{Nortmann} {et~al.}(2018){Nortmann}, {Pall{\'e}}, {Salz},
  {Sanz-Forcada}, {Nagel}, {Alonso-Floriano}, {Czesla}, {Yan}, {Chen},
  {Snellen}, {Zechmeister}, {Schmitt}, {L{\'o}pez-Puertas}, {Casasayas-Barris},
  {Bauer}, {Amado}, {Caballero}, {Dreizler}, {Henning}, {Lamp{\'o}n}, {Montes},
  {Molaverdikhani}, {Quirrenbach}, {Reiners}, {Ribas}, {S{\'a}nchez-L{\'o}pez},
  {Schneider}, \& {Zapatero Osorio}}]{nortmann18}
{Nortmann}, L., {Pall{\'e}}, E., {Salz}, M., {et~al.} 2018, Science, 362, 1388

\bibitem[{Oklop{\v{c}}i{\'c} \& Hirata(2018)}]{oklopcic18}
Oklop{\v{c}}i{\'c}, A., \& Hirata, C.~M. 2018, The Astrophysical Journal
  Letters, 855, L11.
\newblock \url{http://stacks.iop.org/2041-8205/855/i=1/a=L11}

\bibitem[{Osterbrock \& Ferland(2006)}]{osterbrock06}
Osterbrock, D.~E., \& Ferland, G.~J. 2006, Astrophysics Of Gas Nebulae and
  Active Galactic Nuclei (University science books)

\bibitem[{{Pizzolato} {et~al.}(2003){Pizzolato}, {Maggio}, {Micela},
  {Sciortino}, \& {Ventura}}]{pizzolato03}
{Pizzolato}, N., {Maggio}, A., {Micela}, G., {Sciortino}, S., \& {Ventura}, P.
  2003, \aap, 397, 147

\bibitem[{{Ramsey} {et~al.}(1998){Ramsey}, {Adams}, {Barnes}, {Booth},
  {Cornell}, {Fowler}, {Gaffney}, {Glaspey}, {Good}, {Hill}, {Kelton},
  {Krabbendam}, {Long}, {MacQueen}, {Ray}, {Ricklefs}, {Sage}, {Sebring},
  {Spiesman}, \& {Steiner}}]{HETRamsey}
{Ramsey}, L.~W., {Adams}, M.~T., {Barnes}, T.~G., {et~al.} 1998, in \procspie,
  Vol. 3352, Advanced Technology Optical/IR Telescopes VI, ed. L.~M. {Stepp},
  34--42

\bibitem[{{Roberge} \& {Dalgarno}(1982)}]{roberge82}
{Roberge}, W., \& {Dalgarno}, A. 1982, \apj, 255, 489

\bibitem[{SageMath(2019)}]{sage}
SageMath, I. 2019, CoCalc Collaborative Computation Online, {\tt
  https://cocalc.com/}

\bibitem[{{Salz} {et~al.}(2016){Salz}, {Czesla}, {Schneider}, \&
  {Schmitt}}]{salz16}
{Salz}, M., {Czesla}, S., {Schneider}, P.~C., \& {Schmitt}, J.~H.~M.~M. 2016,
  \aap, 586, A75

\bibitem[{{Salz} {et~al.}(2018){Salz}, {Czesla}, {Schneider}, {Nagel},
  {Schmitt}, {Nortmann}, {Alonso-Floriano}, {L{\'o}pez-Puertas}, {Lamp{\'o}n},
  {Bauer}, {Snellen}, {Pall{\'e}}, {Caballero}, {Yan}, {Chen}, {Sanz-Forcada},
  {Amado}, {Quirrenbach}, {Ribas}, {Reiners}, {B{\'e}jar}, {Casasayas-Barris},
  {Cort{\'e}s-Contreras}, {Dreizler}, {Guenther}, {Henning}, {Jeffers},
  {Kaminski}, {K{\"u}rster}, {Lafarga}, {Lara}, {Molaverdikhani}, {Montes},
  {Morales}, {S{\'a}nchez-L{\'o}pez}, {Seifert}, {Zapatero Osorio}, \&
  {Zechmeister}}]{salz18}
{Salz}, M., {Czesla}, S., {Schneider}, P.~C., {et~al.} 2018, \aap, 620, A97

\bibitem[{Seager \& Sasselov(2000)}]{seager00}
Seager, S., \& Sasselov, D.~D. 2000, The Astrophysical Journal, 537, 916.
\newblock \url{http://stacks.iop.org/0004-637X/537/i=2/a=916}

\bibitem[{{Shetrone} {et~al.}(2007){Shetrone}, {Cornell}, {Fowler}, {Gaffney},
  {Laws}, {Mader}, {Mason}, {Odewahn}, {Roman}, {Rostopchin}, {Schneider},
  {Umbarger}, \& {Westfall}}]{HETqueue}
{Shetrone}, M., {Cornell}, M.~E., {Fowler}, J.~R., {et~al.} 2007, \pasp, 119,
  556

\bibitem[{{Spake} {et~al.}(2018){Spake}, {Sing}, {Evans}, {Oklop{\v c}i{\'c}},
  {Bourrier}, {Kreidberg}, {Rackham}, {Irwin}, {Ehrenreich}, {Wyttenbach},
  {Wakeford}, {Zhou}, {Chubb}, {Nikolov}, {Goyal}, {Henry}, {Williamson},
  {Blumenthal}, {Anderson}, {Hellier}, {Charbonneau}, {Udry}, \&
  {Madhusudhan}}]{spake18}
{Spake}, J.~J., {Sing}, D.~K., {Evans}, T.~M., {et~al.} 2018, \nat, 557, 68

\bibitem[{{Stefansson} {et~al.}(2016){Stefansson}, {Hearty}, {Robertson},
  {Mahadevan}, {Anderson}, {Levi}, {Bender}, {Nelson}, {Monson}, {Blank},
  {Halverson}, {Henderson}, {Ramsey}, {Roy}, {Schwab}, \&
  {Terrien}}]{stefansson16}
{Stefansson}, G., {Hearty}, F., {Robertson}, P., {et~al.} 2016, \apj, 833, 175

\bibitem[{Stefansson {et~al.}(2020)Stefansson, Ca{\~{n}}as, Wisniewski,
  Robertson, Mahadevan, Maney, Kanodia, Beard, Bender, Brunt, Clemens, Cochran,
  Diddams, Endl, Ford, Fredrick, Halverson, Hearty, Hebb, Huehnerhoff,
  Jennings, Kaplan, Levi, Lubar, Metcalf, Monson, Morris, Ninan, Nitroy,
  Ramsey, Roy, Schwab, Sigurdsson, Terrien, \& Wright}]{stefansson20}
Stefansson, G., Ca{\~{n}}as, C., Wisniewski, J., {et~al.} 2020, The
  Astronomical Journal, 159, 100.
\newblock \url{https://doi.org/10.3847%2F1538-3881%2Fab5f15}

\bibitem[{{Stone} \& {Proga}(2009)}]{stone09}
{Stone}, J.~M., \& {Proga}, D. 2009, \apj, 694, 205

\bibitem[{Tange(2011)}]{Tange2011a}
Tange, O. 2011, ;login: The USENIX Magazine, 36, 42.
\newblock \url{http://www.gnu.org/s/parallel}

\bibitem[{{Turner} {et~al.}(2016){Turner}, {Christie}, {Arras}, {Johnson}, \&
  {Schmidt}}]{turner16}
{Turner}, J.~D., {Christie}, D., {Arras}, P., {Johnson}, R.~E., \& {Schmidt},
  C. 2016, \mnras, 458, 3880

\bibitem[{van~der Walt {et~al.}(2011)van~der Walt, Colbert, \&
  Varoquaux}]{Numpy2011}
van~der Walt, S., Colbert, S.~C., \& Varoquaux, G. 2011, Computing in Science
  Engineering, 13, 22

\bibitem[{Vidal-Madjar {et~al.}(2003)Vidal-Madjar, des Etangs, Désert,
  Ballester, Ferlet, Hébrard, \& Mayor}]{vidalmadjar03}
Vidal-Madjar, A., des Etangs, A.~L., Désert, J.-M., {et~al.} 2003, Nature,
  422, 143.
\newblock \url{https://www.nature.com/articles/nature01448}

\end{thebibliography}

\end{document}